\documentclass[12pt,a4paper]{article}
\usepackage[toc,page]{appendix}
\usepackage{fullpage}
\usepackage[latin1]{inputenc}
\usepackage[french,english] {babel}
\usepackage[sort&compress,numbers,colon,merge]{natbib}
\usepackage{amsmath}
\usepackage{slashed}
\usepackage{amsfonts}
\usepackage{amssymb}
\usepackage{graphicx}
\usepackage{epstopdf}
\usepackage{color}
\usepackage{listings}
\usepackage[stable]{footmisc}
\usepackage{verbatim}
\usepackage{subfig}
\usepackage[top=2.5cm,bottom=2.1cm,right=2.5cm,left=2.5cm]{geometry}
\usepackage{amsthm}
\usepackage{feynmp}
\usepackage{array}
\usepackage[normalem]{ulem}
\definecolor{vert}{rgb}{0,0.5,0}
\usepackage{hyperref}

\newcommand{\beqn}{\begin{eqnarray}}
\newcommand{\eeqn}{\end{eqnarray}}
\newcommand{\beqas}{\begin{eqnarray*}}
\newcommand{\eeqas}{\end{eqnarray*}}
\newcommand{\beq}{\begin{equation}}
\newcommand{\eeq}{\end{equation}}
\newcommand{\bit}{\begin{itemize}}
\newcommand{\eit}{\end{itemize}}
\newcommand{\bseq}{\begin{subequations}}
\newcommand{\bal}{\begin{align}}
\newcommand{\eal}{\end{align}}
\newcommand{\eseq}{\end{subequations}}
\newcommand{\nn}{\nonumber}
\newcommand{\bs}{\boldsymbol}

\begin{document}
\begin{flushright}
{ULB-TH/15-08}
\end{flushright}
\vskip 1cm
\begin{center}
{\huge Triangle Inequalities for Majorana-Neutrino Magnetic Moments}
\vskip .7cm
{\large Jean-Marie Fr\`ere\footnote{\texttt{frere@ulb.ac.be}}},
{\large Julian Heeck\footnote{\texttt{julian.heeck@ulb.ac.be}}}, and
{\large Simon Mollet\footnote{\texttt{smollet@ulb.ac.be}}}
\vskip .7cm
\emph{Service de Physique Th\'{e}orique,
Universit\'{e} Libre de Bruxelles,\\
Boulevard du Triomphe, CP225, 1050 Brussels, Belgium}
\end{center}

\begin{abstract}
 Electromagnetic properties of neutrinos, if ever observed, could help to decide the Dirac versus Majorana nature of neutrinos. We show that the magnetic moments of Majorana neutrinos have to fulfill triangle inequalities, $|\mu_{\nu_\tau}|^2 \leq |\mu_{\nu_\mu}|^2 +| \mu_{\nu_e}|^2$ and cyclic permutations, which do not hold for Dirac neutrinos. Observing a violation of these inequalities, e.g.~by measuring the magnetic moment of $\nu_\tau$ at SHiP, would thus strongly hint either at the Dirac nature of neutrinos or at the presence of at least one extra light sterile mode.
\end{abstract}

\section{Introduction}

Neutrino oscillations have provided conclusive evidence for non-vanishing neutrino masses by observing non-zero mass-squared differences $m_i^2-m_j^2$~\cite{Giunti:2014ixa}. The absolute mass scale is not known, but bound to be below $\sim$ eV from cosmology and beta-decay measurements. On the theoretical side the problem arises how to include neutrino masses in the very successful Standard Model (SM).
One solution is to introduce $SU(2)_L\times U(1)_Y$ singlet right-handed neutrinos and standard (tiny) Yukawa couplings, making neutrinos Dirac particles just like the charged fermions. The alternative is to allow for lepton number violation by two units, in which case Majorana mass terms for the neutrinos can be written down or arise radiatively. There are a variety of possibilities, the simplest being to include a scalar triplet in the Lagrangian, but the most common are the seesaw mechanism and its variants, in which the neutrino masses are suppressed by a very large right-handed neutrino mass.

The Dirac versus Majorana nature of neutrinos -- indistinguishable with neutrino oscillations -- is of obvious interest to our understanding of particle physics, as it can be linked to the fate of lepton number, and ultimately to the baryon asymmetry of the Universe.
Unfortunately, establishing the neutrino nature is notoriously difficult (see Ref.~\cite{Zralek:1997sa} for a pedagogical review of the issue).
Observation of neutrinoless double beta decay (and, more generically, of any process that violates lepton number by two units) would be direct evidence of the Majorana character~\cite{Schechter:1981bd}, and impressive progress has been made over the last decades to search for this elusive decay mode~\cite{Rodejohann:2011mu}.
Unfortunately, the lack of neutrinoless double beta decay at a given precision level cannot in itself decide in favor of a Dirac character, as the combination of masses, mixing angles, and phases appearing in the decay rate (the so-called \textit{effective Majorana mass} $\langle m_{\beta\beta}\rangle$) may indeed be strongly suppressed (for instance in pseudo-Dirac schemes).
More can be expected once further measurements of the neutrino spectrum (namely the absolute neutrino mass scale and/or hierarchy) will be combined with neutrinoless double beta limits, as the result can in some cases exclude the pure Majorana case (assuming the absence of even more new physics that destructively interferes).

In this note we point out a quite different approach, based on magnetic moments (as we discuss later, we include under ``magnetic'' moments both the electric and magnetic moments). It is well known that Fermi statistics allow a two-component neutrino to possess a mass, but prevent a magnetic moment. Unfortunately (and quite obviously), this argument does not apply to transition magnetic moments involving different flavours.  This observation thus seems to be without effect for the determination of the nature of neutrinos, since actual direct measurements cannot distinguish between diagonal and transition moments. The reason for this is simple: assuming a pure flavour neutrino is produced and scatters on a target through magnetic coupling, there is no way to identify the out-going neutrino, so we measure only an effective sum of couplings.

All hope is not lost, however: the antisymmetric nature of Majorana-neutrino magnetic moments leaves an imprint in the form of inequalities, which we will discuss below.
A small proviso is required here. It is well known that neutrino magnetic moments
are strongly suppressed in the SM (extended by right-handed
neutrinos/Dirac masses), with bounds of the order of $10^{-24} \mu_B$, way below the
current limits (see Sec.~\ref{sec:experiment}). Observation of neutrino magnetic moments in itself (irrespective of
the following considerations) would thus be a proof of physics beyond the SM.
Examples of such physics leading to sizable magnetic moments can be found in
Refs.~\cite{Voloshin:1987qy,Babu:1989wn,Barbieri:1990qj,Barr:1990um} (see also Ref.~\cite{Boyarkin:2014oza} for a recent evaluation in left--right models).
While the techniques for the measurement of magnetic moments (which for terrestrial
experiments  exploit the energy dependence of the neutrino cross sections) goes beyond
the scope of the present paper, it is amusing that it could in some cases even lead
to kinematic zeros~\cite{Bernabeu:2004ay}!

\section{Two-Component Neutrinos and Magnetic Moments}

Currently the best way to probe magnetic moments of neutrinos in terrestrial experiments is to study neutrino scattering off electrons (see Sec.~\ref{sec:experiment}). We begin by considering the simplest case: three light two-component neutrinos, which are simply the ``active'' ones (getting their mass for instance through a scalar triplet), so that no other fermions are involved for the moment.
 Since we are going to discuss realistic measurements of magnetic moments, we choose to formulate the problem first with neutrinos in the flavour basis, and use Weyl (left-handed) neutrinos (in four dimensions Weyl and Majorana are equivalent representations). In four-component spinor notation this means $ \nu_L \equiv \frac{(1 -\gamma_5)}{2}  \nu_L\equiv P_L \nu_L$.
The relevant effective Hamiltonian describing electric and magnetic dipole moments  of neutrinos then reads
\begin{equation}\label{magnMomentWeyl}
    H_\text{eff} = \frac{\mu_{IJ}}{2} \overline{\nu^c_I} \sigma_{\alpha \beta} P_L  \nu_J  F^{\alpha \beta}  + \text{h.c.},
\end{equation}
where $\mu_{IJ}$ are complex numbers, and $ I,J = e,  \mu , \tau$ are flavour indices. (Note that this notation covers both ``electric'' and ``magnetic'' moments.) $F^{\alpha\beta}$ is the field-strength tensor of the photon and $\sigma^{\mu\nu} \equiv \gamma^\mu \gamma^\nu - \gamma^\nu\gamma^\mu$.

Since we only have two-component neutrinos (we will keep calling them Majorana modes, although we use the Weyl notation) the matrix $\mu_{IJ}$ is antisymmetric, so it contains only three free (complex) parameters. Despite the absence of diagonal magnetic moments, experiments are still sensitive to  the transition moments and it will be practically impossible to distinguish them from the diagonal ones, as the final state neutrino can not be observed.

The typical experimental set-up is composed of a production line yielding for instance $\mu$ neutrinos, which later interact with a target. The emerging neutrino $\nu_X$ after the interaction $\nu_\mu e^- \to \nu_X e^-$ will be a linear combination, never to be detected. We thus have the following vertex:
\begin{equation}\label{effective mu }
  \sqrt{|\mu_{e\mu}|^2 + |\mu_{\tau\mu}|^2}
   \left(\overline{\nu^c_X} \sigma_{\alpha \beta} \nu_{\mu} F^{\alpha \beta}\right) ,
\end{equation}
where
\begin{equation}\label{coherent state}
  \overline{ \nu^c_X} \equiv \frac{(\mu_{e\mu} \overline{\nu^c_e}  + \mu_{\tau\mu} \overline{\nu^c_{\tau}}) }{  \sqrt{|\mu_{e\mu}|^2 + |\mu_{\tau\mu}|^2}} \,.
\end{equation}
In this effective coupling, we have properly normalized the out-going neutrino as a (coherent) superposition of flavour states.

The measured result is thus equivalent to an effective magnetic moment (in that it has apparently the same effect as a $\mu_{\mu\mu}$ in scattering):
\begin{equation}\label{apparent mu defined}
    |\mu_{\nu_\mu}| \equiv  \sqrt{|\mu_{e\mu}|^2 + |\mu_{\tau\mu}|^2} \,.
\end{equation}
The same can be done for $\nu_e$ and $\nu_\tau$ scattering off electrons, and it is a simple exercise to check the following inequalities:
\begin{align}
\begin{split}
 |\mu_{\nu_\tau}|^2 &\leq |\mu_{\nu_e}|^2 +| \mu_{\nu_\mu}|^2 \,,\\
 | \mu_{\nu_\mu}|^2 &\leq  |\mu_{\nu_\tau}|^2 +  |\mu_{\nu_e}|^2\,, \\
 |\mu_{\nu_e}|^2 &\leq  |\mu_{\nu_\mu}|^2 +| \mu_{\nu_\tau}|^2 \,,
	\end{split}
	\label{eq:inequalities}
\end{align}
i.e.~$|\mu_{\nu_\tau}|^2 \leq |\mu_{\nu_\mu}|^2 +| \mu_{\nu_e}|^2$ and cyclic permutations of the indices.
The crucial ingredient for these quadratic triangle inequalities is the antisymmetry of $\mu_{IJ}$, which only holds for Majorana neutrinos.
In case one of the three $\mu_{IJ}$ vanishes, one of the above inequalities becomes an equality, e.g.~$\mu^2_{\nu_e}+\mu^2_{\nu_{\mu}} = \mu^2_{\nu_{\tau}}$ for $\mu_{e\mu} \to 0$.

\begin{figure}[t]
\centering
\includegraphics[width=0.6\textwidth]{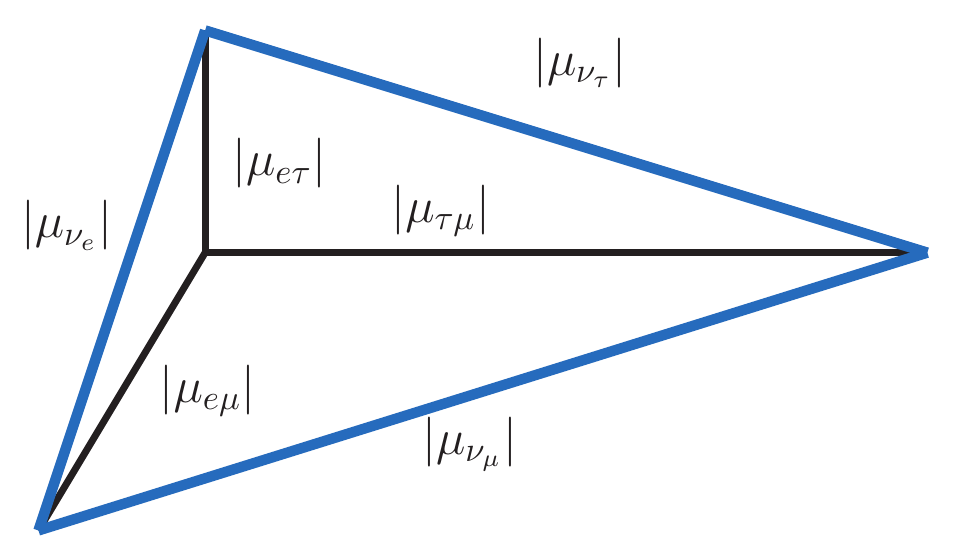}
\caption{
$|\mu_{\nu_J}|$ forms a right triangle with $|\mu_{IJ}|$ and $|\mu_{KJ}|$ (for $I\neq J\neq K$). $|\mu_{\nu_{I,J,K}}|$  thus also form a triangle (shown in thick blue), in general not with right angles.}
\label{fig:magnetic_moment_triangle}
\end{figure}

From the definition of $|\mu_{\nu_{J}}|$ in Eq.~\eqref{apparent mu defined} we see that $|\mu_{\nu_{J}}|$ can be identified as the hypotenuse of a right triangle, with the $|\mu_{IJ}|$ forming the remaining sides (see Fig.~\ref{fig:magnetic_moment_triangle}). This shows that the three $\mu_{\nu_{J}}$ form a triangle as well, although in general not a right one (except when one $\mu_{IJ}$ vanishes). From this we also see that the more generic triangular inequalities also apply:
\begin{equation}
||\mu_{\nu_J}|-|\mu_{\nu_K}| |  \leq |\mu_{\nu_I}| \leq |\mu_{\nu_J}|+|\mu_{\nu_K}|\,,
\end{equation}
for $I\neq J\neq K\neq I$. These inequalities are easier to interpret geometrically, but not as strong as those from Eq.~\eqref{eq:inequalities}, 
as they are automatically satisfied if Eq.~\eqref{eq:inequalities} holds:
\begin{align}
|\mu_{\nu_I}|\leq \sqrt{ |\mu_{\nu_J}|^2+|\mu_{\nu_K}|^2} \leq \sqrt{ |\mu_{\nu_J}|^2+2 |\mu_{\nu_J}| |\mu_{\nu_K}| +|\mu_{\nu_K}|^2} = |\mu_{\nu_J}|+|\mu_{\nu_K}|\,.
\end{align}
The origin of the quadratic equalities of Eq.~\eqref{eq:inequalities} can be traced back to the fact that the $|\mu_{\nu_{I,J,K}}|$ triangle comes from a slice through a tetrahedron with a right-angle corner (Fig.~\ref{fig:magnetic_moment_triangle}), which is not the most general triangle and hence subject to more special constraints.

The experimental falsification of any of the inequalities in Eq.~\eqref{eq:inequalities} would invalidate the underlying hypothesis that the three neutrinos are Majorana particles. One can easily convince oneself that Dirac neutrinos can (but do not have to) violate these inequalities thanks to their diagonal magnetic moments.

The above demonstration (we will discuss the experimental situation in Sec.~\ref{sec:experiment}) rests on the additional implicit hypothesis that the incident neutrinos are indeed three orthogonal flavour eigenstates, in other words that oscillations between the production and interaction do not contaminate significantly the process.

The above formulation is completely general if only three light neutrinos are present. In case of mixing with heavy states, e.g.~in a seesaw scenario, the light modes can still be written by convention as left-handed spinors, possibly after CP conjugation.
The only difference in that case is that the spinors are then no longer pure members of an $SU(2)_L$ doublet. (In a seesaw model, they would include a small part of $\nu_R^c$, the CP conjugate of the right-handed partner, which behaves of course as a left-handed spinor.) As we will discuss explicitly in the next section, the conclusions would not be affected if the oscillations between production and interaction can be neglected, even in this extension.

We insist that the move to a Majorana or mass eigenstate basis is not needed in the above approach. In such a basis, the complex magnetic moment $\mu$ introduced before would be separated into real and imaginary parts, and a distinction between electric and magnetic moment would arise. It is in principle not needed in the above presentation, within the limits described, but we will provide a derivation in the next section.

What would however upset the identities is the presence of additional light modes beyond the three active ones considered here. One trivial example is the Dirac case where three light $\nu_R$ are added and allow for $L$--$R$ magnetic moments. A more ``economical'' case is the presence of even a single (a fortiori, more) light sterile neutrino $\nu_{X,\, L}$, which can induce both fast oscillations (depending on its mass), and transition magnetic moments $\nu_i$--$\nu_X $ which invalidate the stated inequalities (see also Ref.~\cite{Balantekin:2013sda}). Basically, the inequalities rest simply on the number of light neutrino modes (with light Dirac neutrinos counted as two modes).

\section{In a (Majorana) Mass Basis}

The present section derives the above triangle inequalities in the mass basis of neutrinos in a more common notation, given in Ref.~\cite{Giunti:2014ixa}.
It comes as no surprise that the inequalities of Eq.~\eqref{eq:inequalities} hold true after a proper definition of the appropriate effective magnetic moment.
The effective one-photon coupling of neutrinos with the electromagnetic field is described by the electromagnetic current matrix element
\beq
\langle \nu_a\vert j_{\mu}(0)\vert\nu_b\rangle = \bar{u}_a \Lambda_{\mu}^{ab}u_b \,,
\eeq
where $\nu_a,\nu_b$ represent massive neutrino states and $u_a, u_b$ their spinors.
Hermiticity and gauge invariance strongly restrict the form of the matrix $\bs{\Lambda}_{\mu}$, which is a matrix both in spinor and family space:
\beq
\bs{\Lambda}_{\mu}= (\gamma_{\mu}-q_{\mu}\slashed{q}/q^2) \left[\bs{F_Q}+\bs{F_A} q^2 \gamma_5\right]
-i\sigma_{\mu\nu}q^{\nu} \left[\bs{F_M}+i\bs{F_E}\gamma_5 \right],
\eeq
where $\bs{F_Q}, \bs{F_M}, \bs{F_E}$, and $\bs{F_A}$ are hermitian matrices depending only on $q^2$ (the photon momentum), and which correspond to charge, magnetic dipole, electric dipole, and anapole form factors, respectively. For $q^2=0$ (real photon), these correspond to the charge ($\bs{q}$), the magnetic moment ($\bs{\hat{\mu}}$), the electric moment ($\bs{\hat{\epsilon}}$) and the anapole moment ($\bs{a}$), respectively.\footnote{The hats are used to differentiate between the hermitian moments $\hat{\mu}$ and $\hat{\epsilon}$ and the complex moment $\mu$ defined in the previous section (which includes both electric and magnetic contributions).}

The Majorana character of neutrinos imposes additional constraints on the form factors: $\bs{q},\bs{\hat{\mu}}$ and $\bs{\hat{\epsilon}}$ are antisymmetric, while $\bs{a}$ is symmetric. Note that the antisymmetric property of $\bs{\hat{\mu}}$ and $\bs{\hat{\epsilon}}$ is independent of the basis in flavour space. Mathematically, this shows why a single Majorana neutrino cannot possess (among other things) a magnetic dipole moment.

If we focus on the dipole form factors ($\bs{\hat{\mu}}$ and $\bs{\hat{\epsilon}}$), it is easy to show that this corresponds to the effective Hamiltonian (\ref{magnMomentWeyl}) in the mass basis with the definition $\mu=\hat{\mu}+i\hat{\epsilon}$.

Now let us come back to our realistic experiment where a neutrino $\nu_I$ with a certain flavour $I$ is created and then propagates towards a detector where it interacts with an electron through its magnetic and electric dipole moment. It is converted into a massive neutrino $\nu_b$ with an amplitude
\beq
A_{Ib}\propto \sum_a U^*_{Ia}e^{-iE_aT+i\bs{p\cdot L}}(\hat{\mu}_{ab}-i\hat{\epsilon}_{ab}) \,.
\eeq
Again, we cannot observe the out-going neutrino and thus have to sum over all mass eigenstates in the cross section:
\beq
\sigma\propto \sum_b\left| \sum_a U^*_{Ia}e^{-i\Delta m^2_{ab}L/2E}(\hat{\mu}_{ab}-i\hat{\epsilon}_{ab})\right| ^2\equiv (\mu^2_I)_{\text{eff}} \,,
\eeq
which defines the effective magnetic moment $(\mu^2_I)_{\text{eff}}$ for the scattering $\nu_I e^- \to \nu e^-$. In experiments with a short baseline $L$ -- compared to $E_\nu/\Delta m_{ij}^2$, $E_\nu$ being the neutrino energy -- we can neglect propagation just as in the previous section. Using the antisymmetry properties of Majorana neutrinos we find
\beq
(\mu^2_I)_{\text{eff}}\simeq \sum_a(\tilde{\mu}^2_a+\tilde{\epsilon}^2_a)-\left|\sum_a U_{Ia}(\tilde{\mu}_a-i\tilde{\epsilon}_a)\right|^2,
\eeq
where we have defined $\hat{\mu}(\hat{\epsilon})_{ab}=i\sum_c \varepsilon_{abc}\tilde{\mu}(\tilde{\epsilon})_c$~\cite{Giunti:2014ixa}.
For three different neutrino flavours $I,J,K$ ($I\neq J$, $J\neq K$, $K\neq I$), we have
\beqn
(\mu^2_I)_{\text{eff}}+(\mu^2_J)_{\text{eff}}-(\mu^2_K)_{\text{eff}}
&=& \sum_a(\tilde{\mu}^2_a+\tilde{\epsilon}^2_a)-\sum_{a,b} \left(\delta_{ab}-2U_{Ka}U^*_{Kb}\right)
(\tilde{\mu}_a-i\tilde{\epsilon}_a)(\tilde{\mu}_b+i\tilde{\epsilon}_b)\nn\\
&=& 2\left|\sum_a U_{Ia}(\tilde{\mu}_a-i\tilde{\epsilon}_a)\right|^2 \geq 0\,,
\eeqn
thanks to the unitarity of $U$, which is an alternate proof of the inequalities in Eq.~\eqref{eq:inequalities}.

\section{Experimental status}
\label{sec:experiment}

The arguably cleanest experimental probe for electromagnetic properties is neutrino scattering off of electrons, with the electron's recoil and scattering angle as handles to distinguish the new physics from SM contributions.
The electron-neutrino's effective magnetic moment is constrained by GEMMA~\cite{Beda:2012zz} (reactor $\bar\nu_e$--$e^-$), the muon neutrino's by LSND~\cite{Auerbach:2001wg} (accelerator $\nu_\mu,\bar\nu_\mu$--$e^-$), and the $\tau$ neutrino's by DONUT~\cite{Schwienhorst:2001sj} (accelerator $\nu_\tau,\bar\nu_\tau$--$e^-$). At $90\%$~C.L., the limits are
\begin{align}
|\mu_{\nu_e}| < 2.9\times 10^{-11} \mu_B\,, &&
|\mu_{\nu_\mu}| < 6.8\times 10^{-10} \mu_B\,, &&
|\mu_{\nu_\tau}| < 3.9\times 10^{-7} \mu_B\,.
\end{align}
GEMMA-II (in preparation) is expected to have a sensitivity down to $\mu_{\nu_e}\sim 1\times 10^{-11}\mu_B$~\cite{Beda:2013mta}, while the recently planned
COHERENT experiment at the Spallation Neutron Source can improve the limit on $\mu_{\nu_\mu}$ by half an order of magnitude~\cite{Kosmas:2015sqa}. To check whether the inequalities of Eq.~\eqref{eq:inequalities} are valid, the most important magnetic moment is however $\mu_{\nu_\tau}$, as it is the least constrained one.
Fortunately, the proposed SHiP (Search for Hidden Particles) experiment at CERN SPS~\cite{Anelli:2015pba} is expected to study $\mathcal{O}(4000)$ $\nu_\tau$ interactions, compared to the handful of $\nu_\tau$ observed so far in DONUT and OPERA. With the number of expected events $N_\mathrm{ev} = 4.3\times 10^{15} (|\mu_{\nu_\tau}|/\mu_B)^2$, we can hope for an improvement on $\mu_{\nu_\tau}$ of more than an order of magnitude, the actual number being subject to a full detector specification and simulation~\cite{Alekhin:2015byh}.
SHiP is hence in a position to strongly hint towards the Dirac nature of neutrinos simply by observing the magnetic moment of $\nu_\tau$.

There are caveats to the above discussion, because magnetic moments are not just constrained using man-made neutrino beams. Using instead the natural neutrino sources of the Sun and atmosphere, one typically finds constraints of order $10^{-10}\mu_B$ on certain effective neutrino magnetic moments (either squared sum or weighted with mixing angles)~\cite{Giunti:2014ixa}.
Even more stringent bounds come from astrophysical data, not based on neutrino--electron scattering.
A magnetic-moment enhanced plasmon decay rate $\gamma\to \bar\nu \nu$ will for example drastically change the energy-loss rate of stars. A study of the red-giant branch in globular clusters allows putting a $95\%$~C.L.~limit of $4.5\times 10^{-12}\mu_B$ on $\mu_\nu$ (which is effectively a sum over all $\mu_{\nu_\ell}$)~\cite{Viaux:2013lha}.

The astrophysical limits are not only orders of magnitude stronger than laboratory bounds, but also fairly insensitive to the different neutrino flavour or mass eigenstates, and cannot be used on their own to check our inequalities.
The strong astrophysical bounds  should, however, not prevent the search for neutrino magnetic moments in terrestrial experiments.
As an example, astrophysical measurements rely on very low momentum behaviour (the conversion of keV photons), and would
not be sensitive to a magnetic transition between $\nu_i$--$\nu_X$ if the sterile mass is larger than this energy. In a similar way, terrestrial experiments have various energy (and hence mass gap) sensitivities, which should be kept in mind.

A possible observation of $\mu_{\nu_\tau}$ in the near future will hence also have a large impact on our understanding of astrophysics.

\section{Conclusions}

The observation of neutrino masses in oscillation experiments provides stunning evidence for physics beyond the Standard Model. While electromagnetic properties of neutrinos are typically tiny in minimally extended models, there exist many models with potentially observable couplings~\cite{Voloshin:1987qy,Babu:1989wn,Barbieri:1990qj,Barr:1990um,Boyarkin:2014oza}. In this letter we pointed out the existence of triangle inequalities among the effective magnetic moments
\beq
|\mu_{\nu_{\tau}}|^2 \leq |\mu_{\nu_{\mu}}|^2+ |\mu_{\nu_e}|^2\nn \quad \text{ and cyclic permutations,}
\eeq
which, however, only hold if neutrinos are of Majorana nature (by which we mean that they have only two degrees of freedom, whether they are written in Majorana or Weyl notation).
These inequalities constitute basically a count of the number of light degrees of freedom (three two-component fermions for the Majorana case), and their violation indicates the presence of extra light modes. These could be the extra $\nu_R$ included in Dirac neutrinos, or additional light (sterile) neutrinos. More exotic scenarios could involve new forces which violate lepton flavor universality, e.g.~new (light) neutral gauge bosons $Z'$, which would, however, lead to a different momentum dependence in neutrino scattering.
Practically, the most promising approach to checking these inequalities is to improve the limits on $\mu_{\nu_\tau}$, which is possible at the proposed SHiP experiment at CERN SPS.

\section*{Acknowledgements}

This work is funded in part by IISN and by Belgian Science Policy (IAP VII/37 ``Fundamental Interactions'').

\bibliographystyle{utcaps_mod}
\bibliography{BIB}

\end{document}